# Memory Offloading for Large Language Model Inference with Latency SLO Guarantees

Chenxiang Ma
Peking University

Zhisheng Ye
Peking University

Hanyu Zhao
Alibaba Cloud Computing

Zehua Yang
Peking University

Tianhao Fu
Peking University

Jiaxun Han
Peking University

Jie Zhang
Peking University

Yingwei Luo
Peking University

Xiaolin Wang
Peking University

Zhenlin Wang
Michigan Tech

Yong Li
Alibaba Cloud Computing

Diyu Zhou
Peking University

## Abstract

Offloading large language models (LLMs) state to host memory during inference promises to reduce operational costs by supporting larger models, longer inputs, and larger batch sizes. However, the design of existing memory offloading mechanisms does not take latency service-level objectives (SLOs) into consideration. As a result, they either lead to frequent SLO violations or underutilize host memory, thereby incurring economic loss and thus defeating the purpose of memory offloading.

This paper presents SELECT-N, a latency-SLO-aware memory offloading system for LLM serving. A key challenge in designing SELECT-N is to reconcile the tension between meeting SLOs and maximizing host memory usage. SELECT-N overcomes it by exploiting a unique characteristic of modern LLMs: during serving, the computation time of each decoder layer is deterministic. Leveraging this, SELECT-N introduces offloading interval, an internal tunable knob that captures the tradeoff between SLOs and host memory usage, thereby reducing the aforementioned challenge to pick an optimal offloading interval. With that, SELECT-N proposes a two-stage approach to automatically pick the offloading interval. The first stage is offline that generates the range of optimal offloading interval, while the second stage adjusts offloading interval at the granularity of inference iteration based on runtime hardware status. Our evaluation shows that SELECT-N consistently meets SLOs and improves the serving throughput over existing mechanisms by 1.85× due to maximizing the use of host memory.

## 1 Introduction

Large language models (LLMs), a recent advancement in deep learning, have revolutionized multiple computing domains, including content generation [3, 45, 70], data analysis [31, 48, 50], language translation [20, 40, 49], and code development [11, 27, 41, 65], thanks to their ability to perform various complex tasks with remarkable accuracy.

However, a key challenge to democratizing LLM is to meet its immense memory demand during inference. Modern LLMs often come with tens to hundreds of billions of parameters with serving requests often involving long prompts and/or large batch sizes. The memory demand is further exacerbated by emerging deployment scenarios, like multi-round conversation [24] and long-form text generation[10]. As a result, serving modern LLMs typically demands hundreds of gigabytes of memory, far exceeding the memory capacity of even advanced GPUs like NVIDIA A100 [14]. To meet the memory demands, the current industry practice involves deploying a model on multiple GPUs to use their aggregate memory. However, this approach incurs significant operational costs.

A promising approach [12] to meet the memory demands is to offload part of the LLM state to host memory (*i.e.*, those used by CPU) during serving and transfer the offloaded state back to GPU when needed. State-of-the-art offloading mechanisms are DeepSpeed Inference (DeepSpeed afterwards) [8] and FlexGen [58]. Both mechanisms leverage the fact that modern LLM serving performs computation on one decoder layer (layer afterwards) at a time before moving to the next layer. Specifically, DeepSpeed only keeps the minimal necessary LLM state in GPU memory, namely the state of the current layer, while offloading all other LLM states to host memory. FlexGen, instead, offloads a fixed portion of the state across all layers to the host memory. This portion is determined by factors like model size, batch size, and sequence length, and the decision is made statically using metrics like peak GPU performance, without executing the serving requests on the GPU. To mitigate the performance impact incurred by offloading, both systems prefetch the state of the next layer during the computation of the current layer.

A key limitation of both DeepSpeed and FlexGen is that their design does not consider an important factor: the request serving must meet certain latency service level objectives (SLOs). The latency SLOs exist because many LLM deployment scenarios, such as chatbots [17, 22] and virtual assistants [60], involve frequent interaction of users. In this case, the serving must be highly responsive, and any single SLO violation incurs severe economic losses [52], since this harms the user experience, increasing the likelihood of users switching to competing services.

As a result, both DeepSpeed and FlexGen become highly ineffective when latency SLOs must be considered. Deep-

Speed significantly increases the serving time (9.5× as shown in Figure 2 (a)), resulting in frequent SLO violations. This limitation is due to the design choice of only keeping the current layer in GPU memory, and if the layer transfer time exceeds its computation time, which is often the case with modern LLMs, the serving latency significantly increases. Unfortunately, as discussed above, SLO violations incur a significant economic loss, defeating the purpose of offloading to reduce operational costs.

On the other hand, while one can fine-tune input factors in FlexGen to meet latency SLOs, its static decision process forces users to account for the worst case by offloading a theoretical lower bound of memory to the host. A direct consequence of such memory underutilization is that FlexGen only allows for a smaller batch size than necessary, leading to a 1.85× drop in throughput. We identified two key factors which, when estimated statically, contribute to memory underutilization: 1) use the peak GPU performance to estimate the model execution time; and 2) upon PCIe bandwidth contention, where $n$ GPUs sharing a single bus, assuming each GPU always only gets $1/n$ of the memory bandwidth.

This paper presents SELECT-N, an effective latency-SLO-aware memory offloading system for LLM inference. The overarching design goal of SELECT-N is to maximize host memory usage while guaranteeing the latency SLOs. To meet this goal, the design of SELECT-N exploits a unique characteristic of LLM serving: the computation time on each layer is deterministic (*i.e.*, always the same). This is because each layer consists of the same structure and always performs the same set of operations with the input of identical size.

At the core of SELECT-N is offloading interval, an internal tunable knob that effectively captures the tension between meeting SLOs and maximizing host memory usage. An offloading interval of $i$, means, during serving, for every $i$ layer, SELECT-N offloads the last layer of the interval to the host memory. Importantly, unlike DeepSpeed and FlexGen which only prefetch the offloaded state when the computation reaches the previous layer, SELECT-N prefetches the offloaded layer much earlier, when it starts to compute the first layer in the offloading interval. This enables SELECT-N to use a larger offloading interval to handle stricter SLOs since it keeps more state in GPU, thereby using additional computation to cover state transfer latency. On the other hand, with a looser SLO, SELECT-N uses a small offloading interval to maximize host memory usage. Crucially, the trade-off can be made by adjusting the offloading interval because the deterministic execution time of LLM layers ensures that the computation time for all layers, whether within the same iteration or across different iterations, remains the same.

Following the above setup, meeting the design goals of SELECT-N is simplified to choosing an optimal offloading interval, *i.e.*, the smallest possible one that meets the SLO, for each serving request. SELECT-N achieves this in two steps and, unlike FlexGen, performs them dynamically based on the actual execution. The first step decides the optimal offloading interval assuming no PCIe bandwidth contention. This step is performed once for each new LLM deployed and is conducted on a dedicated offline server. In this step, unlike FlexGen, which estimates the serving time based on the GPU's peak performance, SELECT-N again leverages the deterministic execution time of the LLM to execute a stream of prompts on the model to generate a performance record. The record stores, for each valid combination of target SLOs, batch sizes, and sequence lengths, the optimal offloading interval. Therefore, upon receiving a request, SELECT-N directly obtains the optimal offloading interval from the record, avoiding any extra latency from online measurements.

Finally, to handle PCIe bandwidth contention, the second step is performed on the actual servers that handle user requests. In this case, a per-bus coordinator constantly monitors the bandwidth utilization of all GPU sharing the PCIe bus and, based on the offloading interval obtained from the first step, adjusts the offloading interval of each GPU accordingly to ensure each does not violate SLOs, while maximizing the total host memory usage. Our extensive experiments demonstrate that, unlike DeepSpeed, SELECT-N can always correctly maintain latency SLOs across a range of setups. In addition, given the same latency SLOs, SELECT-N uses 2.37× more host memory than FlexGen, leading to a throughput increase of 1.85×.

In summary, we make the following contributions.
- **Offloading interval.** We propose offloading interval, a simple yet effective abstraction that captures the tension between serving latency and memory saving, leveraging a unique characteristic of LLM: the computation time of each layer is deterministic.
- **Offline performance analyzing.** We propose an effective offline performance analysis technique, again leveraging the determinism in LLM computation time.
- **SELECT-N.** We design and implement SELECT-N, a complete and, to our knowledge, the first latency-SLO-aware memory offloading system for LLM serving.

We are committed to open-sourcing SELECT-N.

## 2 Background

This section presents the necessary background to facilitate the discussion for the rest of the paper, including a primer on large language model (LLM) inference (§2.1), the need to maintain latency SLOs during LLM inference (§2.2), and memory offloading techniques for LLM (§2.3).

### 2.1 A Primer on LLM Inference

Large language models (LLMs), a new class of deep learning models, have demonstrated great success in various domains, such as data analysis [31, 48, 50], language translation [20, 40, 49], content generation [3, 45, 70], and code development [11, 27, 41, 65]. Such success is due to the fact that LLM can easily generalize across different tasks and efficiently process large amounts of unstructured data [53].

During inference, an LLM takes as input a sequence of tokens ( e.g., typically an English word), which is called a prompt. The sequence length denotes the number of tokens in a prompt. In addition, to maximize LLM inference performance, a common approach is to batch multiple prompts together to allow the GPU to process them simultaneously. The batch size is the number of prompts in a batch. The output of an LLM, called an output sequence, is also a sequence of tokens.

This paper focuses on decoder-only LLMs, such as GPT [55], LLaMA [64], and OPT [76], which are arguably the most common type of LLMs. These models are based on the Transformer [66] architecture, consisting of multiple decoder layers (layers afterwards). Each layer has the same structure with the same number of matrices, with each corresponding matrix having the same dimensions. The only difference is the values of the elements within each matrix.

The computation during LLM inference is conducted in a layer-by-layer fashion; LLM performs computation on one layer, updates the relevant state, and then moves to the next layer. Each layer performs the same set of operations, which involve attention computations to capture dependencies between tokens and MLP computations to transform and refine the representations.

**Prefill and Decoding.** LLM inference consists of two phases: prefill and decoding. During prefill, an LLM traverses each layer to generate the first token in the output sequence by processing all the tokens in the input sequence. During decoding, the model generates the rest of the tokens iteratively. In the first iteration, the model uses the token generated by the prefill phase to produce the next token in the output sequence. Afterwards, the output token generated in the previous iteration is used as the input for the next iteration. The process stops when reaching a word limit or seeing a set of predefined end tokens or end patterns.

Critically, the prefill and decoding exhibit different characteristics. The prefill phase is computation-intensive since it involves processing all the tokens in an input sequence. The decoding phase is memory-intensive and only needs to process the token generated in the previous iteration.

### 2.2 LLM Inference with SLO requirements

As discussed earlier, due to their powerfulness, LLMs are used for various tasks. Some of these tasks have a long serving time, on the order of a few hours, and do not require real-time interactions with human users. Examples include summarizing or translating large text corpora [7, 23, 80], bulk content generation [39], or offline analytics [72, 79].

On the other hand, lots of LLM tasks, such as chatbots [17, 22] and virtual assistants [60], involve frequent interactions with human users. Thus, these tasks must be highly responsive and are often associated with stringent latency service level objectives (SLOs) that must be met since failing to do so incurs a severe economic loss. For exam-ple, with production chatbot services, one latency SLO is that each token must be generated within hundreds of milliseconds to ensure timely feedback that aligns with human reading speeds [26]. Failing to meet the SLO would cause the users to experience delays and thus be more likely to switch to competing services.

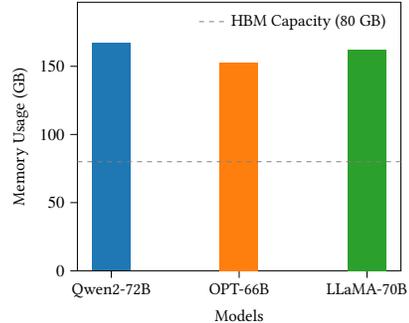

**Figure 1:** Memory demands of modern LLMs using float16 precision, with input sequences of 2,048 tokens. The grey dashed line represents the GPU memory capacity (80GB) of the NVIDIA A100.

This paper targets two common latency SLOs: (1) Time to First Token (TTFT) (*i.e.*, the amout of time to generate the first token of a response); and (2) Time Per Output Token (TPOT) (*i.e.*, the average amount of time to produce the rest of the token, excluding the first one). TTFT and TPOT are used for SLOs of the prefill and decoding phases, respectively.

### 2.3 Offloading LLM to Host Memory

Modern LLMs often have tens to hundreds of billions of parameters with long prompt lengths. As a result, these models require hundreds of gigabytes of memory to handle daily requests, as shown in Figure 1 This demand far exceeds the capacity of even cutting-edge GPUs like the NVIDIA A100 [14], which only equips 80GB of memory.

To address the growing memory demands of LLMs, a promising approach [12] is to offload part of the model state to host memory, which is traditionally used by CPUs. The offloaded state is transferred back to the GPU as needed. Offloading effectively reduces GPU memory usage, thereby achieving various benefits, including 1) enabling the deployment of powerful models that exceed GPU memory capacity, 2) allowing for the generation of longer output sequences, and 3) supporting larger inputs with greater batch sizes and/or input sequences.

This offloading is effective because host memory is inherently larger than GPU memory. Modern servers often have several terabytes of host memory, while the memory of modern GPUs is often only a few tens of gigabytes. The orders-of-magnitude difference is due to 1) GPU memory, built using high-speed HBM and GDDR technologies, is significantly more expensive than the DDR SDRAM used in host memory, and 2) extending GPU memory is more challenging than host memory due to constraints on power consumption and heat dissipation.

## 3 Motivation

This section presents prior memory offloading approaches for LLM serving (§3.1), and explains why they are insufficient to handle requests associated with SLOs (§3.2, §3.3, §3.4).

Unless otherwise mentioned, we conduct all the experiments in this section with an NVIDIA A10 GPU with 24GB memory. The PCIe bandwidth is 24GB/s. We use TPOT and TTFT as latency SLOs (§2.2) The more detailed experimental setup is presented in §5.1.

### 3.1 Prior Offloading Approaches

There have been only a few prior mechanisms for offloading LLM state to CPU memory during inference. Among them, the most relevant ones are DeepSpeed Inference (DeepSpeed afterwards) [8] and FlexGen [58], while others are orthogonal to SELECT-N, as discussed in §6.

DeepSpeed and FlexGen work as follows. DeepSpeed keeps only the state of the layer currently being computed in GPU memory while offloading all other states to host memory. Unlike DeepSpeed, FlexGen offloads a fixed portion of the state (referred to as the offloaded portion) for all layers to host memory. FlexGen selects an offloaded portion that maximizes inference throughput and decides the exact offloaded portion upon receiving a serving request. The decision is made by solving a linear programming problem, taking as input various factors, such as model size, batch size, sequence length of the serving request, and the transfer speed between the host and the GPU memory.

Naturally, offloading is prone to performance overhead, where the computation may need to wait for the offloaded state to be ready in the GPU memory. Both DeepSpeed and FlexGen mitigate this overhead by overlapping data transfer with computation: they prefetch the offloaded state of the next layer while computing the current layer.

However, the design of DeepSpeed and FlexGen does not consider an important aspect: the latency SLO of the serving requests. Such a limitation is rooted in their design. As a result, as we elaborate next, DeepSpeed suffers from frequent SLO violations, while FlexGen has to estimate for the worst case to avoid SLO violations, thereby underutilizing host memory and failing to achieve optimal throughput.

### 3.2 Keeping Minimal LLM State in GPU

With DeepSpeed, SLO violations are due to its key design choice that only keeps the minimal absolutely necessary LLM state in GPU memory (i.e., the state of the current layer). This design maximizes host memory usage, but the inference performance heavily hinges on that the computation of the current layer can cover the transfer time of the next layer.

Unfortunately, for most modern LLMs, the computation is much shorter than the transfer time of a layer. We demonstrate this in Figure 2 (b), where the transfer time is 3.5× and 13.8× longer than the computation time for the prefill and decoding phase, respectively. As a result, as shown in Figure 2 (a), DeepSpeed increases the serving latency by up to 9.5×, resulting in frequent SLO violations for all evaluated models. In addition to latency, since the GPU waits for the data most of the time, such an approach also reduces throughput by up to 8.2×, as shown in Figure 3.

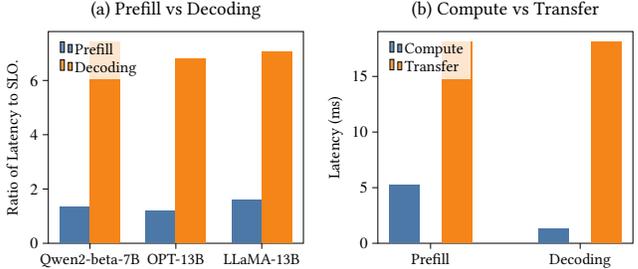

**Figure 2:** (a) Serving latency (normalized by the target SLO) with DeepSpeed. (b) The average computation and transfer time for a single layer. Model: Qwen2-beta-7B, sequence length: 256, batch size: 4.

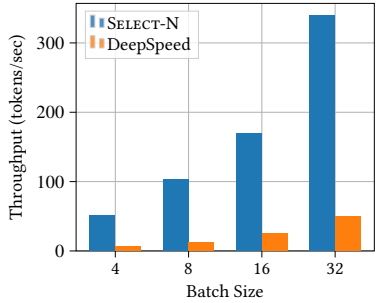

**Figure 3:** Throughput of SELECT-N and DeepSpeed with varying batch sizes. Model: Qwen2-beta-7B

This is a major limitation since, for many user-facing LLM tasks, such as those discussed in §2.2, meeting the latency SLO is the top priority. Even a single SLO violation incurs a severe economic loss, thus defeating the purpose of minimizing operational costs in offloading approaches. Therefore, DeepSpeed is limited to only those LLM tasks that do not require human interactions.

**Obvervation #1: Keeping only one layer in GPU, as done by DeepSpeed, is prone to severe SLO violations.**

### 3.3 Estimating Execution Time

As discussed in §3.1, the optimization goal of FlexGen is to maximize serving throughput. To evaluate FlexGen in our target scenario, we made a slight modification to the decision algorithm in FlexGen: it now takes a target SLO as input and output maximum offloaded portion. We verify that this modified, SLO-aware version of FlexGen functions correctly and can successfully meet the specified SLOs. For simplicity, we uses FlexGen to refer to this modified version throughout the rest of the paper.

We found that a fundamental limitation of FlexGen is that it decides the offloaded portion (§3.1) statically (before the requests are actually executed); in other words, FlexGen cannot adjust the offloaded portion during execution based on current system status. Therefore, to avoid SLO violations, with FlexGen, one has to conservatively estimate for the worst case, making FlexGen underutilize host memory but rather uses much more GPU memory than necessary. As a result, FlexGen often fails to achieve the optimal performance, as shown in Figure 5. The SLO is the TPOT when running without offloading. In this case, FlexGen uses 2.1× less host memory than SELECT-N, and thus supports smaller batch sizes, leading to 1.9× throughput reduction. We note that this is the best case for FlexGen and a more relaxed SLO makes the throughput difference even larger.

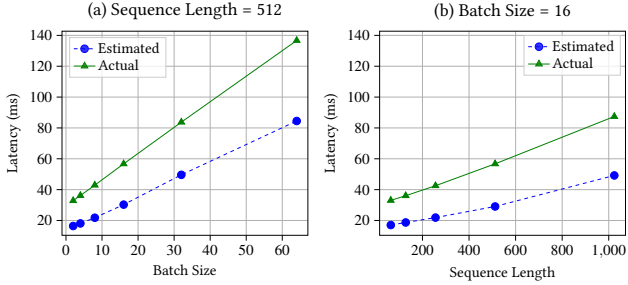

**Figure 4:** The actual serving latency vs. the one estimated by FlexGen. Model: OPT-13B.

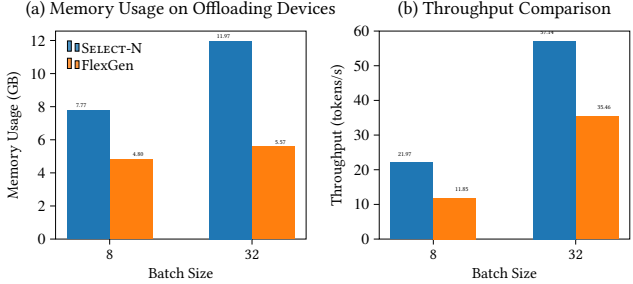

**Figure 5:** Comparison of SELECT-N and FlexGen in (a) Memory usage on the offloading devices and (b) Throughput. Model: OPT-13B.

With FlexGen, an important factor for underutilizing host memory is that FlexGen statically estimates serving latency using peak GPU performance. This is necessary to avoid SLO violations since the estimated value is the theoretical lower bound of the serving latency. However, as shown in Figure 4, the estimated latency is much shorter than the real one. This makes FlexGen offload a smaller amount of model state than is necessary to host memory, since FlexGen mistakenly believes that the computation time is not long enough to overlap the transfer time of the larger model state.

**Obvervation #2: Estimating model execution time using peak GPU performance, as with DeepSpeed, is inaccurate and heavily underutilizes host memory.**

### 3.4 Bandwidth Contention

Another important factor for FlexGen to underutilize host memory is the contention on transfer bandwidth between host and GPU memory. Specifically, to minimize operational costs, the current industry practice is to place multiple GPUs on a single machine node. However, some of these GPUs share a single PCIe bus (which connects the host and GPUs), and thus contends for the PCIe bandwidth [44, 59]. As a result, the available bandwidth to transfer LLM state for each individual GPU fluctuates during serving a request.

With FlexGen, in the presence of bandwidth contention, to meet SLO, one must again estimate the worst case: each GPU only gets $1/n$ of the bandwidth, where $n$ is the number of GPUs that share a PCIe bus. However, for a given GPU, the available bandwidth may fluctuate but is often much larger than the worst-case one, since other GPUs may be idle, or do not fully utilize their share of the bandwidth. As a result, FlexGen overestimates the transfer time, again offloading a smaller amount of LLM state than is necessary to host memory.

**Observation #3: FlexGen statically estimates the PCIe bandwidth under contention, thereby underutilizing the host memory.**

## 4 The SELECT-N System

This section presents SELECT-N, an latency-SLO-aware memory offloading system for LLM inference. This section starts with SELECT-N's design goals (§4.1), an overview (§4.2), followed by the design of each component.

### 4.1 Design Goals.

We design SELECT-N with the following goals.
- **Meeting latency SLOs.** Departing from DeepSpeed, the top priority of SELECT-N is to meet latency SLOs, aligning with the overarching goal of memory offloading: reducing operational costs.
- **Maximizing host memory usage.** Once adhering to latency SLO, unlike FlexGen, SELECT-N should maximize host memory usage to e.g., support larger models, enable greater batch sizes and/or longer sequence lengths, or allow the models to produce longer outputs.
- **Fine-grained dynamic adjusting.** SELECT-N should dynamically (rather than statically as with FlexGen) decide the amount of memory offloaded to the host, considering factors such as the sequence length and batch size of serving requests, the associated latency SLOs and, particularly, the current machine status.

### 4.2 Overview

**Deployment scenario.** SELECT-N operates on a single machine equipped with multiple GPUs, where each hosts a model. These GPUs may contend on the PCIe bandwidth. SELECT-N takes as input a serving request and its associated SLO. Such an SLO may not be the end-to-end one: upper level components can adjust the SLO passed to SELECT-N based on, e.g., networking delays that are already incurred. SELECT-N next checks whether this SLO can be met by the GPUs it manages (§4.5), as there are situations where the SLO cannot be met at all. For example, a deployed model with weights requiring memory far exceeding the GPU's capacity forces a large state to be offloaded to host memory, causing long data transfer times that violate the SLO. If the SLO can be met, SELECT-N schedules the request on one of the GPUs. If not, SELECT-N passes the request to the upper-level scheduler, which can avoid the SLO violation, by, *e.g.*, sending the requests to other node hosting models without memory offloading.

**Components and workflow.** To meet the design goals, at its core, SELECT-N operates on the offloading interval (§4.3), an internal tunable knob that captures the tradeoff between meeting SLOs and maximizing host memory usage. A small offloading interval makes SELECT-N offload more LLM state to host memory and thus may potentially slow down inference, being more prone to SLO violations. A large offloading interval achieves the opposite. As further explained in §4.3, using offloading interval to control the aforementioned tradeoff is enabled by a unique characteristic of LLM: during serving, each layer takes the same amount of computation time. Thanks to offloading interval, meeting the design goals of SELECT-N is reduced to automatically and dynamically adjusting the offloading interval for each GPU instance. SELECT-N achieves this in two stages: first an offline stage and then an online stage, as we detail next.

As shown in Figure 6, SELECT-N consists of three components: 1) a performance analyzer, to find the optimal offloading interval (*i.e.*, the smallest one that meets SLO) under no bandwidth contention; 2) per-GPU runtime memory managers, which take as input offloading interval, and transfers layer state between GPU and host memory basd on the offloading interval; and 3) per-bus runtime bandwidth coordinators, which adjust the offloading interval for all GPUs sharing a bus at the granularity of each inference iteration.

The performance analyzer operates offline (*i.e.*, not on servers that handle user requests) to find the optimal offloading interval. Specifically, to ensure no bandwidth contention, the performance analyzer operates on a dedicated server, where each GPU exclusively occupies a single PCIe bus. Upon deploying a new model on a GPU that SELECT-N manages, the model is passed to the performance analyzer. The analyzer generates a stream of prompts and executes them on the model to generate a performance record. A performance record stores the optimal offloading interval for all valid combinations of SLOs, sequence length, and batch size. Generating a performance record beforehand is possible due to, again, LLM's deterministic execution time. Unlike performance analyzer, the memory manager and bandwidth coordinators operate online on normal servers that serve user requests.

The workflow of SELECT-N is as follows. When a request comes, ❶ SELECT-N first waits for a GPU instance hosting the corresponding model to become available. ❷ Based on the target SLO (minus the waiting time), the sequence length, and the batch sizes of the request, SELECT-N consults the performance record of the model to obtain the optimal of-floading interval. ❸ The offloading interval and the requests are passed to the bandwidth coordinator, which generates an adjusted offloading interval for each GPU instance that shares the bus considering their bandwidth utilization. ❹ The bandwidth coordinator passes the request to the selected GPU, and the set of adjusted offloading interval to the runtime memory managers of all GPUs sharing the bus. The runtime memory manages applies the adjusted offloading interval before the next inference iteration.

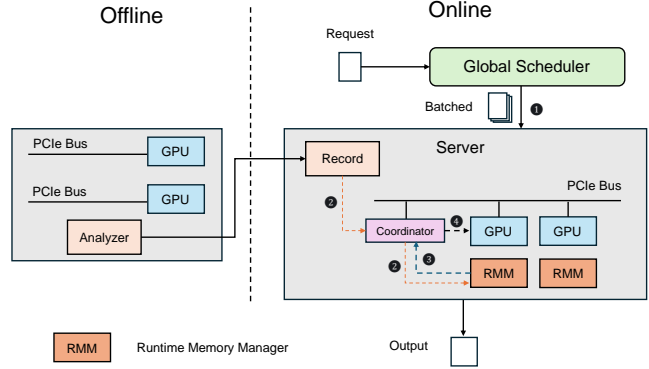

**Figure 6:** The architecture and workflow of SELECT-N.

### 4.3 Runtime Memory Manager

**Insight.** A key design behind offloading interval is to exploit a special characteristic of LLM: during inference, the computation time of each layer is always identical, even in different iterations. This is because, as discussed in §2.1, 1) each layer consists of the same structure (*i.e.*, same number of matrices of the same size for the corresponding matrices); 2) each layer performs the same operations, and 3) the size of the input to each layer remains the same (*i.e.*, either all input tokens in prefill or one token in decoding).

**Offloading interval.** Using this, SELECT-N proposes offloading interval. An offloading interval of $i$ means that for every $i$ layer, the state of the last layer is offloaded to CPU memory, while the state of other layers are always in GPU memory. We term the last layer offloaded layer.

The memory manager in SELECT-N transfers the state between host and GPU memory following a decided offloading interval. To maximize performance, SELECT-N also follows the design scheme of overlapping computation with data transfer (§4.2). However, unlike DeepSpeed and FlexGen that prefetches the offloaded layer only when the computation is on the exact previous layer, the memory manager in SELECT-N prefetches the state of the offloaded layer by initiating the loading upon the computation on the first layer in the offloading interval. Therefore, with SELECT-N, the transfer time is hidden by the computation time of multiple layers, rather than one layer as DeepSpeed and FlexGen.

Figure 7 concretizes the above discussion by showing a scenario where the offloading interval is 4. In this case, the states of layers 1 to 3 and 5 to 7 are always in the GPU memory, while layers 4 and 8 are offloaded layer. Before computing layers 1 and 5, Select-N prefetches the state of layers 4 and 8, respectively. Next, before computing layers 4 and 8, Select-N ensures that their state is ready in the GPU memory. Once the computation is done, Select-N moves their state back to CPU.

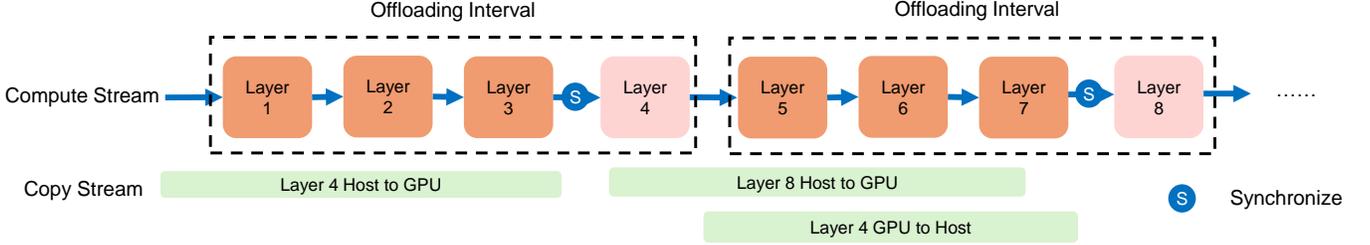

**Figure 7:** An overview of how a memory manager works. The upper part represents the compute stream, where each block denotes a layer with pink blocks being the offloaded layers. The lower part is the copy stream. The two streams execute in parallel, with "S" meaning synchronization points between the streams.

The offloading interval offers an effective mechanism to resolve the tension between meeting the SLO and maximizing host memory usage. A small offloading interval maximizes host memory usage, since more layers are offloaded there, but can only support looser SLO, since there are fewer layers whose computations are used to hide the load of the offloaded layer. In fact, Select-N is reduced to DeepSpeed if the offloading interval is set to 1. A larger offloading interval increases GPU memory usages, but can meet stricter SLO.

The above design simplifies resolving the aforementioned tension to selecting an optimal (*i.e.*, the smallest offloading interval that meets the SLO of given input requests). The following subsections present how Select-N performs the selection with a two-phase approach.

### 4.4 Generating Performance Record

**Challenge.** To decide the optimal offloading interval, one must know the computation time and the transfer time of each layer. A challenge Select-N encounters is that while the computation time of each layer is highly deterministic (*i.e.*, the time is always the same), it is not predictable; one cannot easily estimate beforehand the computation time of each layer, as we have shown in §3.3. As a result, this forces Select-N to measure this computation time during runtime. An naive approach is to measure the computation time on the real production server every time the actual requests arrive. However, this 1) increases GPU usage, thereby leading to extra operational costs; 2) incurs an extra latency caused by measurements during runtime.

**Insight.** Select-N overcomes this challenge, again by leveraging the deterministic nature of the computation time on each layer. Our observation is that, with the deterministic nature, for a given model and hardware platform, assuming no bandwidth contention, the computation time of each layer depends solely on the size of the prompts, namely the sequence length and batch size. Therefore, Select-N can obtain the accurate execution time of each layer by simply

|    | 128 | 256 | 512 | 1024 |
|----|-----|-----|-----|------|
| 4  | 5   | 4   | 3   | 2    |
| 8  | 4   | 3   | 2   | 1    |
| 16 | 3   | 2   | 1   | 1    |
| 32 | 2   | 1   | 1   | 1    |
| 64 | 1   | 1   | 1   | 1    |

**Table 1:** Performance record for a given SLO. Row: batch size. Column: sequence length. Larger batch sizes and sequence lengths are omitted, as the optimal offloading interval is one.

executing one iteration of the model on the target machine.

**Design.** The above insight allowed Select-N to decide the optimal offloading interval offline with a model analyzer. The analyzer is invoked every time a new model is scheduled to be deployed on a GPU instance managed by Select-N. The analyzer takes as input a list of potentially possible SLOs, and for each SLO, generates a performance record. The performance record stores, for a pair of input sequences and batch size, the optimal offloading interval, assuming no bandwidth contention. Table 1 shows one such example.

Under the hood, the analyzer works in the following steps. For a given SLO, it first enumerates all possible pairs of batch size and sequence length. For any given pair, the analyzer generates a prompt of that size with random content. It next uses the prompt to run one iteration of prefill and decoding. During execution, the analyzer measures 1) $t_{\text{trans}}$: the time to transfer a layer from CPU to GPU memory; 2) and then $t_{\text{compute}}$: the computation time of a single layer. Therefore, to ensure the layer transferring time does not cause SLO violations, the maximum number of layers that can be offloaded is $L_{\text{offload}} = \left\lfloor \frac{t_{\text{compute}} \cdot (1+\delta)}{t_{\text{trans}}} \right\rfloor$, where $\delta$ is the SLO quotient over the computation time without offloading. Finally, the optimal offloading interval for this pair of sequence length and batch size is given by $\left\lfloor \frac{L}{L_{\text{offload}}} \right\rfloor$, where $L$ is the total number of decoder layers in the model.

Creating performance records is practical for the following reasons. First, the number of combinations of batch size, sequence length, and SLOs that need to be enumerated is actually small, since the analyzer only needs to operate at a specific granularity for these values to be effective. For example, the current prototype uses 2-millisecond granularity for SLOs and requires batch sizes and sequence lengths to be powers of two. Select-N selects the optimal offloading interval for combinations not listed in the performance record by conservatively choosing from the nearest combination. With this granularity, the number of target SLOs in production is typically at the scale of hundreds, since latency SLOs for interactive LLM tasks, which sys targets, rarely exceed one second. Although there can be an infinite number of batch size and sequence length pairs, if their product exceeds a certain threshold, the optimal offloading interval becomes 1, since the computational time of a single layer exceeds the transfer time of that layer. Therefore, the possible pairs of the batch size and sequence length the analyzer needs to sample is at most 100 for our evaluated models. Second, given a batch size and sequence length pair, obtaining the optimal offloading interval is fast, which usually takes 1-2 seconds for our models. The above two factors combined mean that the process of creating a performance record is fast, requiring at most 40 minutes for our models. This is much shorter than the frequency of deploying a new model in production, which is at the scale of months.

**Supporting the separation of prefill and decoding.** An emerging LLM deployment scheme is to deploy prefill and decoding phases on separate instances [51, 78]. Such separation maximizes performance by allowing each stage to fully leverage GPUs optimized for their specific workloads. In addition, the separation facilitates independent scaling of the two phases.

Unfortunately, both DeepSpeed and FlexGen do not consider the separation of prefill and decoding; they offload the same portion of the model state to host memory for both phases, thereby causing significantly longer delays during the decoding phase.

A key benefit enabled by the analyzer is to effectively support the seperation of prefill and decoding. In this case, the analyzer creates the performance record for prefill and decoding separately, thereby considering the different characteristics of the two phases. Hence, departing from prior work, Select-N can use a different offloading interval that is most suitable for the corresponding phase. In general, the compute-intensive prefill phase requires a shorter offloading interval, while the memory-intensive decoding phase requires a larger offloading interval.

### 4.5 Addressing Bandwidth Contention

The analyzer chooses the optimal offloading interval by assuming an ideal scenario where each GPU can utilize the whole PCIe bandwidth. However, as discussed in §3.4, real-world inference scenarios may incur bandwidth contention among different GPU instances. Select-N addresses this issue with a per-bus runtime bandwidth coordinator, that adjusts offloading interval for contented GPU instances at the granularity of each inference iteration (*i.e.*, at the granularity of generating an output token), as we detail next.

The coordinator performs the adjustment by observing that, given an input request, each GPU instance has a minimum and maximum offloading interval. The minimum offloading interval is derived from the performance record (§4.4); any offloading interval below the minimum violates the SLO of the input. The maximum offloading interval depends on the model memory deployed on the GPU. Since Select-N enables deploying models that require more memory than the GPU has, the offloading interval must remain below a threshold to avoid exceeding GPU capacity. For models whose required memory is within the GPU capacity, the maximum offloading interval is infinite.

Critically, the offloading interval also precisely controls the usage of PCIe bandwidth. A small offloading interval incurs higher bandwidth usage since layers are swapped more frequently between host and GPU memory, while a large offloading interval incurs lower bandwidth. Furthermore, given an offloading interval, the consumed bandwidth can be accurately estimated, as shown in Lines 4-13 of Figure 8.

Therefore, with the above setup, to avoid SLO violations, the coordinator needs 1) to pick, for each contented GPU instance, a valid offloading interval that falls between the minimum and maximum interval; and 2) ensures that the sum of the PCIe bandwidth consumed for each offloading interval is below the PCIe bandwidth. In addition, another goal of Select-N is to optimize the right set of offloading interval to maximize the total usage of host memory. Select-N can also adopt other reasonable optimization goals. For the sake of simplicity, the rest of the discussion assumes two contented GPU instances, but the algorithm can be easily extended to more instances.

Figure 8 shows how the coordinator works. Upon receiving a new request, the coordinator first checks if this request's SLO can possibly be met by checking if its minimum offloading interval is less than the maximum one (Lines 34-35). If no valid interval can be found, the request is returned to the upper level scheduler (§4.2). Otherwise, the coordinator enumerates all possible combinations of offloading intervals for the two GPU instances to find all the valid ones (Lines 39-42). Next, the coordinator finds the interval pairs that maximize host GPU memory usage (Line 45). Once the valid interval pairs are decided, the request is served on one GPU with the chosen interval while the other GPU adjusts the interval in the next iteration (Lines 47-48). We note that a special case is that the other GPU is idle, and in this case, the request is served with the minimum offloading interval.

### 4.6 Implementation

Select-N is implemented based on the vLLM framework, leveraging its efficient memory management and inference capabilities. To optimize performance, Select-N uses separate CUDA streams for computation and data transmission. By enabling parallel execution through stream overlap, Select-N significantly reduces latency and improves

```python
def is_valid(intv_1, intv_2, mdl_1, mdl_2):
    # Calculate memory transfer for the first model instance
    mem_transfer_1 = (mdl_1.layers *
        mdl_1.mem_per_layer) / intv_1

    transfer_rate_1 = mem_transfer_1 / mdl_1.iter_time

    # Calculate memory transfer for the second model instance
    mem_transfer_2 = (mdl_2.layers *
        mdl_2.mem_per_layer) / intv_2

    transfer_rate_2 = mem_transfer_2 / mdl_2.iter_time

    # Check if memory transfer rate is within bandwidth
    return transfer_rate_1 + transfer_rate_2 < bandwidth:

def calc_host_memory(intv_1, intv_2, mdl_1, mdl_2):
    # Calculate offloaded layers and memory usage for mdl_1
    layers_offloaded_1 = mdl_1.layers / intv_1
    memory_usage_1 = layers_offloaded_1 * mdl_1.mem_per_layer

    # Calculate offloaded layers and memory usage for mdl_2
    layers_offloaded_2 = mdl_2.layers / intv_2
    memory_usage_2 = layers_offloaded_2 * mdl_2.mem_per_layer

    # Return the total host memory usage
    return memory_usage_1 + memory_usage_2

# req is scheduled to be served by mdl_1
def adjustment_mechanism(req, mdl_1, mdl_2):
    # Check if the SLO of the request can be met
    if req.min_intv > mdl_1.max_intv:
        return False

    # Enumerate all valid interval combinations
    comb = []
    for intv_1 in mdl_1.valid_intervals:
        for intv_2 in mdl_2.valid_intervals:
            if is_valid(intv_1, intv_2, mdl_1, mdl_2):
                comb.append(intv_1, intv_2)

    # Find the pair that maximizes host memory usage
    best_comb = max(comb, key=calc_host_memory)

    mdl_1.adjust(best_comb.intv_1)
    mdl_2.adjust(best_comb.intv_2)

    return True
```

**Figure 8:** The adjustment algorithm performed by the coordinator.

throughput during inference.

We encapsulated SELECT-N into a Python library that allows seamless integration with Transformer-based models. This library dynamically manages the decoder layers of the model, enabling efficient scheduling and resource allocation. With this design, users can easily wrap existing Transformer models to take advantage of SELECT-N without requiring extensive modifications while benefiting from enhanced performance and scalability.

## 5 Evaluation

### 5.1 Experimental Setup

**Model and System Configuration.** The experiments are conducted on a system equipped with 4 NVIDIA A10 GPUs. The models used in this study include OPT-6.7B, OPT-13B, Qwen2-beta-7B, and Llama2-13B. These models were selected to represent a range of sizes and complexities, ensuring a thorough evaluation of the SELECT-N mechanism on varying model scales. Unless otherwise specified, all experiments are conducted using a separation of the prefill and decoding phases, with each model deployed as two distinct instances: a prefill instance and a decoding instance.

**Workload.** The workload used in the experiments is generated using a randomly designed dataloader, which allows users to customize batch size, sequence length, and vocabulary size based on the model requirements. Our setup can equally support real-world workloads, such as ShareGPT[2] and Alpaca[63, 67] datasets, as the fundamental workload characteristics are consistent and do not significantly impact experimental results.

**Baseline.** We consider three baselines in our evaluation: DeepSpeed, FlexGen, and a naive method. The naive method, built upon the vLLM framework, involves loading the entire model into GPU memory without any offloading. Our SELECT-N mechanism is also implemented on top of the vLLM framework, whose paged attention mechanism enables fine-grained memory management during inference.

**Key Metrics.** We use several key metrics to evaluate performance, including GPU memory savings for memory efficiency, TTFT and TPOT for latency, and throughput for overall inference performance across prefill and decoding phases.

### 5.2 Maintaining SLO

We conduct experiments using the OPT-6.7B model and the Qwen2-beta-7B model with clear separation of the prefill and decoding phases to show the capability of SELECT-N in maintaining SLO. The prefill phase corresponds to the TTFT SLO, while the decoding phase corresponds to the TPOT SLO. This separation enables us to apply different offloading interval values to the two instances, allowing each phase to meet its respective SLO effectively.

Due to the benefits of separation, which allows for an increased batch size in the decoding instance as we discussed earlier, we fix the batch size at 128 for the decoding instance and 32 for the prefill instance to simulate a realistic high-throughput inference scenario. As a baseline, we measure the performance under a naive execution mode without offloading, where the TTFT and TPOT latencies are recorded as reference points. Subsequently, we configured SELECT-N to operate under varying SLO constraints for both TTFT and TPOT to test its ability to meet these requirements by adjusting the offloading interval parameter. Furthermore, we conducted experiments to compare the performance of SELECT-N with DeepSpeed.

To ensure consistency, the SLO values are normalized to 1, and the recorded values represent the ratio of the observed time to the specified SLO. The results shown in Figure 9 demonstrate that SELECT-N effectively maintains the specified SLOs for both TTFT and TPOT in different setups. It further illustrates how SELECT-N adapts to varying SLO con-straints by dynamically adjusting the offloading interval parameter. When the SLO is set within the range 20%-40%, the optimal offloading interval remains unchanged, resulting in relatively stable TPOT values. As the SLO increases from 40% to 50%, the optimal offloading interval selected during the experiment changes, causing a noticeable variation in the TPOT curve.

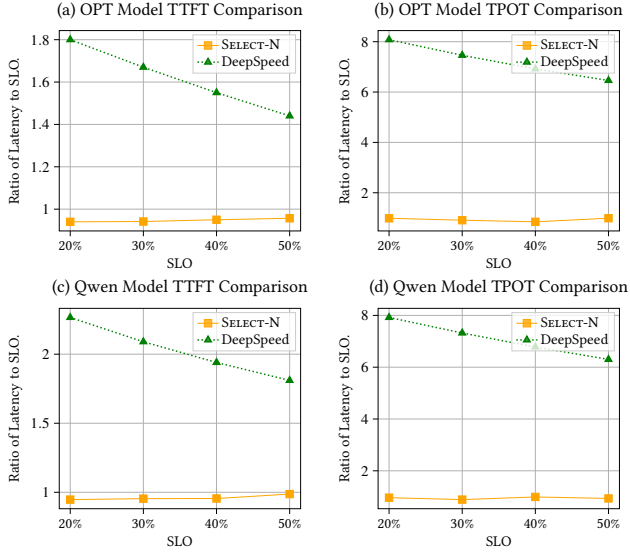

**Figure 9:** Comparison of TTFT and TPOT between SELECT-N and DeepSpeed under the OPT-6.7B and Qwen2-beta-7B models. The y-axis represents the ratio of the observed latency to the corresponding SLO target latency, where a value of 1 indicates that the latency matches the SLO target.

DeepSpeed suffers from significant transmission latency during inference. This bottleneck is particularly severe in the decoding phase, where the large volume of parameter transfers drastically amplifies the delay, resulting in exceeding the specified SLOs by 8.08× and reducing its throughput by 6.8× to 8.23× compared to SELECT-N.

### 5.3 Memory Saving

We conduct experiments to evaluate and compare SELECT-N and FlexGen in terms of memory savings and throughput performance. The experiments use the OPT-13B model across varying batch sizes 4, 8, 16, 32. For different input scales, SELECT-N dynamically adjusts the offloading interval value, while FlexGen relies on its cost model to compute the offload ratio in an attempt to minimize total latency. The comparison focuses on the ability of the two systems to optimize GPU memory usage and maintain high throughput under various input conditions.

Figure 10 presents the results of the comparison. FlexGen's memory-saving capability is consistently inferior to that of SELECT-N at the same batch size due to inaccuracies in its estimation of transfer and computation latencies, resulting in suboptimal offloading decisions. In contrast, SELECT-N employs its analyzer to directly measure computation and data transfer times, allowing it to select the most suitable offloading interval and achieve 2.37× better memory savings compared to FlexGen, thereby maximizing memory efficiency across different input scales.

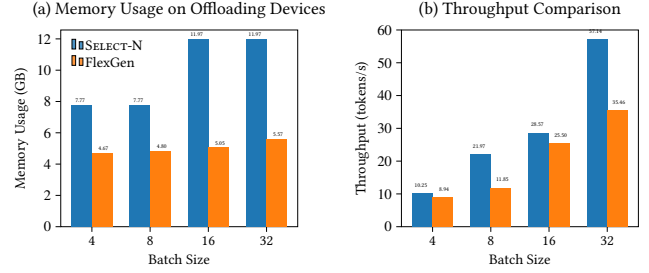

**Figure 10:** (a) Memory usage on the offloading devices for SELECT-N and FlexGen under different batch sizes. (b) Throughput comparison between SELECT-N and FlexGen under different batch sizes.

Furthermore, SELECT-N consistently achieves higher throughput than FlexGen while simultaneously delivering significantly greater memory savings. In the best case, SELECT-N achieves up to 1.85× the throughput of FlexGen. By saving more memory, SELECT-N can support larger input scales, which further enhances throughput. This advantage will be discussed in greater detail in Section §5.6.

### 5.4 Profiling Accuracy

This subsection evaluates the effectiveness of the offloading interval analyzer within the SELECT-N framework, focusing on verifying whether the offloading interval value identified during the record-generating phase is indeed optimal. To validate this process, we perform experiments using the OPT-6.7B model with a clear separation between the prefill and decoding phases. This separation enables the decoding instance to adopt a larger batch size, enhancing throughput in the decoding phase. In our experiments, the sequence length is fixed at 64, with a batch size of 16 for the prefill instance and 128 for the decoding instance. We set the SLO target at 50% and use records from the analyzer to determine the optimal offloading interval value. Subsequently, we evaluated the system's performance under non-optimal offloading interval configurations and recorded the corresponding GPU memory usage for each setting.

The results, shown in Figure 11, highlight the trade-offs inherent in the offloading interval configuration and the accuracy and necessity of the analyzer in the SELECT-N framework. By generating performance records, the optimal offloading interval values are identified as 3 for the prefill phase and 8 for the decoding phase. The optimal offloading interval values achieve an effective balance by ensuring compliance with both TTFT and TPOT SLOs while minimizing GPU memory usage. When offloading interval is smaller than the optimal value, GPU memory usage is reduced; however, this comes at the cost of SLO violations due to increased latency, particularly in the TPOT phase, resulting in degraded throughput. In contrast, larger offloading interval values consistently satisfy the SLO but incur significant memory overhead without measurable performance gains. This inefficiency is evident in Figure 11(c), where increasing offloading interval results in a proportionate consumption of GPU memory while failing to improve latency or throughput.

## 5.5 Bandwidth contention

We conduct experiments simulating a scenario where two GPUs share PCIe bandwidth. In this setup, one GPU runs the OPT-13B model while the other GPU runs the LLaMA-13B model simultaneously. Since both models are too large to fully fit into GPU memory, offloading strategies were required to enable the models to run successfully. That creates a high-bandwidth contention environment as both GPUs perform offloading and data transfers concurrently. The sequence length was set to 64, and batch sizes of 8, 16, and 32 were tested for both tasks.

To address the contention, Select-N dynamically adjusts the offloading interval values for both tasks, ensuring that neither exceeded the predefined SLO. Since it is not feasible to test the computation time in a naive mode where the entire model fits into GPU memory, we could not define the SLO as a percentage of naive execution time. Instead, we set the SLO as a fixed TPOT threshold of 100ms. This value is significantly higher than typical human reading speeds and provides a stringent and practical target for evaluating performance under high-bandwidth contention. We also compare the performance of Select-N with FlexGen on the GPU running the OPT-13B task with the same batch sizes. This comparison highlights Select-N's ability to adaptively manage contention on the shared PCIe bandwidth, compared to FlexGen's static offloading strategy.

The results, shown in Figure 12, illustrate the TPOT performance of Select-N and FlexGen under varying batch sizes in a bandwidth contention scenario. FlexGen exhibits acceptable performance at the largest batch size tested (32) but violates the SLO at smaller batch sizes (8 and 16) due to its inability to adapt to varying PCIe transfer demands. In contrast, Select-N consistently maintains TPOT below the SLO across all batch sizes by dynamically adjusting the offloading interval values of the two tasks, effectively mitigating bandwidth contention and ensuring balanced resource utilization. This adaptability enables Select-N to achieve 2.9× higher throughput compared to FlexGen at smaller batch sizes on the OPT-13B task.

## 5.6 Benefits of Select-N

**Supporting larger models.** Select-N is capable of supporting models whose memory demands exceed the GPU memory capacity, as demonstrated in experiments with the LLaMA-13B and OPT-13B models. Figure 13 presents the TPOT performance of LLaMA-13B and OPT-13B, respectively, under varying batch sizes. Both models, which require memory beyond the 24GB GPU capacity in our setup, were successfully executed using Select-N. TPOT values below 100 ms are higher than the normal human reading speed, indicating that such latencies allow for efficient and real-time text generation. In our experiments, the TPOT values for both LLaMA-13B and OPT-13B remained consistently below 100ms across all tested batch sizes, confirming that Select-N not only enables the execution of these large-scale models but also ensures efficient performance suitable for real-world applications.

**Supporting more input and output tokens.** To evaluate the capability of Select-N in generating longer output sequences and supporting larger batch sizes and/or sequence lengths, we leverage a critical metric: the maximum allocatable length (*max length*). This metric is computed as

$$max\ length = batch\ size \times (sequence\ length + output\ length),$$

where the term captures the total number of tokens that the system can handle for a single model instance. The *max length* is directly determined by the number of GPU blocks allocatable via the vLLM backend, which dynamically manages GPU memory to optimize allocation. A higher *max length* indicates the system's ability to support larger batch sizes, longer input sequences, and extended output sequences.

In our experiments, we use the Qwen2-beta-7B model, which supports a maximum position embedding size of 32,768 tokens. This choice is deliberate, as Qwen2-beta-7B significantly exceeds the position embedding limits of other models like OPT and LLaMA, ensuring that the system remains capable of processing long input sequences without being constrained by the model's internal architecture.

The results in Figure 14 show that, by adjusting the offloading intervals, Select-N achieves varying degrees of memory savings for model inference. When the interval value is smaller, a larger proportion of the model's parameters is offloaded to CPU memory, significantly reducing the GPU memory footprint. This allows Select-N to allocate more GPU memory for token processing, thereby increasing the maximum allocatable length. As a result, the system can support larger batch sizes, longer input sequences, or extended output sequences under smaller interval settings, which helps to improve throughput.

## 6 Related Work

**Efficient LLM serving.** Several recent studies address these challenges by proposing methods to enhance system performance and resource efficiency in LLM inference. Orca[73] introduces continuous batching to enhance GPU throughput. vLLM[42] leverages PageAttention to optimize KV cache memory usage, enabling efficient resource allocation. SARATHI[5] adopts a chunked-prefill strategy, dividing prefill requests into smaller chunks while combining them with decoding requests to achieve better hardware utilization. StreamingLLM[69] extends LLM capabilities by allowing the generation of sequence lengths beyond their original training limits. Select-N builds on some of these techniques like vLLM, and is designed to work in parallel with other approaches to further enhance performance and resource efficiency.

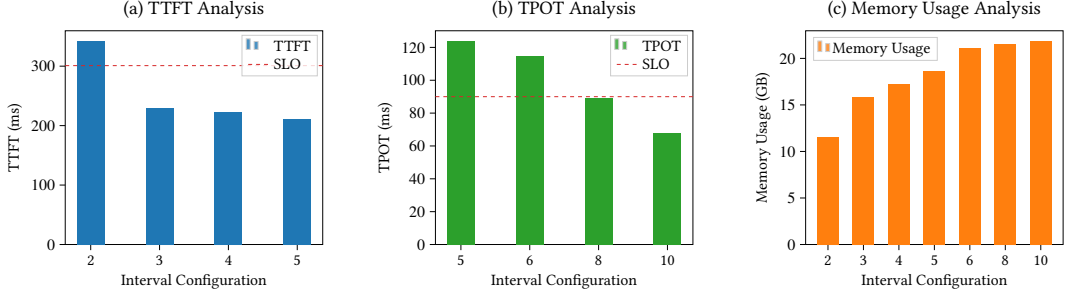

**Figure 11:** The TTFT, TPOT, and memory usage of Select-N under different offloading interval configurations. The red dashed lines represent the SLO. The optimal offloading interval is 3 in (a) and 8 in (b).

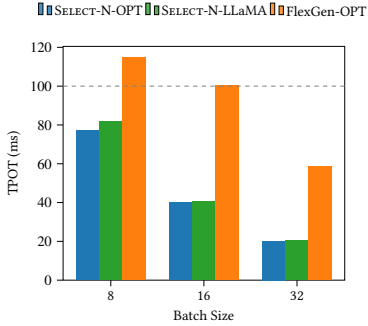

**Figure 12:** TPOT comparison of Select-N (OPT-13B and LLaMA-13B models) and FlexGen (OPT-13B model) under contention environments across different batch sizes. The dashed line represents the SLO.

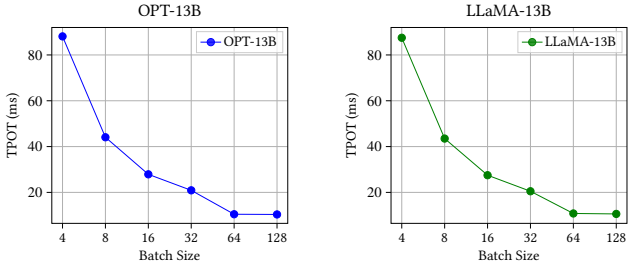

**Figure 13:** TPOT of OPT-13B and LLaMA-13B models using Select-N.

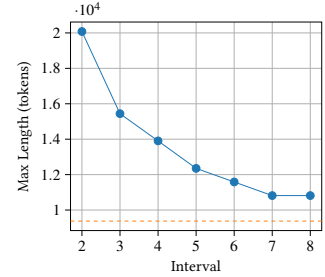

**Figure 14:** Maximum prompt length the model can process under different offloading interval settings. The dashed line represents the maximum length in the naive mode.

**Offloading techniques.** Existing works have explored various techniques to improve large-scale model inference performance, particularly on resource-constrained hardware. Systems such as DeepSpeed ZeRO-Inference[8] and Hugging Face Accelerate[1] adopt offloading strategies originally designed for training scenarios. Infinite-LLM[47] manages the utilization of all GPU and CPU memory resources to store the KV cache. These approaches may still cause computation to stall as they do not ensure data readiness at the required time. InfiniGen[43] mitigates KV cache fetch overhead by speculatively prefetching essential KV entries, improving cache management efficiency. Neo[36] offloads part of attention compute and KV cache states from GPU to CPU to balance compute and memory resources. These two works cannot handle models that exceed the GPU memory capacity, making them orthogonal to our approach.

**Scheduling systems.** Recent works have explored efficient resource scheduling and allocation strategies for deep learning tasks, focusing on optimizing throughput[54], heterogeneous-aware scheduling[35], preemption and latency-aware scheduling[28, 75], and improving resource utilization through model parallelism[46] or iteration-level preemptive scheduling to mitigate queueing delays[68]. There are also concurrent works that employ disaggregation techniques to decouple and balance resource allocation, improving efficiency in LLM inference, such as Splitwise[51], TetriInfer[32], DéjàVu[61], and Distserve[78] Select-N is orthogonal to the large body of work on scheduling, as its separation of prefill and decoding stages can be implemented using any of the aforementioned approaches.

## 7 Conclusion

This paper presents Select-N, a memory offloading mechanism for LLM serving that meets latency SLOs while maximizing the host memory usage. Select-N captures the trade-off between meeting SLOs and maximizing host memory usage with offloading interval, an internal tunable knob. In addition, Select-N automatically decides the optimal offloading interval, *i.e.*, the smallest offloading interval that meets SLOs, with a two-stage tuning approach. The first stage assumes bandwidth contention and profiles the GPU model offline, and generates a performance record that, for any valid combination of SLOs, sequence lengths, and batching sizes, stores an optimal offloading interval, The second stage adjusts the offloading interval for GPU instances sharing the bus to ensure that the SLOs can still be met while maximizing the aggregate host memory usage across all GPU instances. Our evaluation shows that Select-N consistently maintains SLO under various runtime scenarios, and outperforms Flex-Gen in throughput by 1.85×, due to use 2.37× more host memory.

from the bottom up. In *14th USENIX Symposium on Operating Systems Design and Implementation (OSDI 20)*, pages 443–462. USENIX Association, Nov. 2020. ISBN 978-1-939133-19-9. URL https://www.usenix.org/conference/osdi20/presentation/gujarati.

[30] C. Holmes, M. Tanaka, M. Wyatt, A. A. Awan, J. Rasley, S. Rajbhandari, R. Y. Aminabadi, H. Qin, A. Bakhtiari, L. Kurilenko, and Y. He. Deepspeed-fastgen: High-throughput text generation for llms via mii and deepspeed-inference, 2024. URL https://arxiv.org/abs/2401.08671.

[31] S. Hong, Y. Lin, B. Liu, B. Liu, B. Wu, C. Zhang, C. Wei, D. Li, J. Chen, J. Zhang, J. Wang, L. Zhang, L. Zhang, M. Yang, M. Zhuge, T. Guo, T. Zhou, W. Tao, X. Tang, X. Lu, X. Zheng, X. Liang, Y. Fei, Y. Cheng, Z. Gou, Z. Xu, and C. Wu. Data interpreter: An llm agent for data science, 2024. URL https://arxiv.org/abs/2402.18679.

[32] C. Hu, H. Huang, L. Xu, X. Chen, J. Xu, S. Chen, H. Feng, C. Wang, S. Wang, Y. Bao, N. Sun, and Y. Shan. Inference without interference: Disaggregate llm inference for mixed downstream workloads, 2024. URL https://arxiv.org/abs/2401.11181.

[33] C. Hu, H. Huang, L. Xu, X. Chen, J. Xu, S. Chen, H. Feng, C. Wang, S. Wang, Y. Bao, N. Sun, and Y. Shan. Inference without interference: Disaggregate llm inference for mixed downstream workloads, 2024. URL https://arxiv.org/abs/2401.11181.

[34] E. J. Hu, Y. Shen, P. Wallis, Z. Allen-Zhu, Y. Li, S. Wang, L. Wang, and W. Chen. Lora: Low-rank adaptation of large language models, 2021. URL https://arxiv.org/abs/2106.09685.

[35] S. Jayaram Subramanya, D. Arfeen, S. Lin, A. Qiao, Z. Jia, and G. R. Ganger. Sia: Heterogeneity-aware, goodput-optimized ml-cluster scheduling. In *Proceedings of the 29th Symposium on Operating Systems Principles*, SOSP '23, page 642–657, New York, NY, USA, 2023. Association for Computing Machinery. ISBN 9798400702297. doi: 10.1145/3600006.3613175. URL https://doi.org/10.1145/3600006.3613175.

[36] X. Jiang, Y. Zhou, S. Cao, I. Stoica, and M. Yu. Neo: Saving gpu memory crisis with cpu offloading for online llm inference, 2024. URL https://arxiv.org/abs/2411.01142.

[37] H. Jin, R. Lai, C. F. Ruan, Y. Wang, T. C. Mowry, X. Miao, Z. Jia, and T. Chen. A system for microserving of llms, 2024. URL https://arxiv.org/abs/2412.12488.

[38] Y. Jin, T. Wang, H. Lin, M. Song, P. Li, Y. Ma, Y. Shan, Z. Yuan, C. Li, Y. Sun, T. Wu, X. Chu, R. Huan, L. Ma, X. You, W. Zhou, Y. Ye, W. Liu, X. Xu, Y. Zhang, T. Dong, J. Zhu, Z. Wang, X. Ju, J. Song, H. Cheng, X. Li, J. Ding, H. Guo, and Z. Zhang. P/d-serve: Serving disaggregated large language model at scale, 2024. URL https://arxiv.org/abs/2408.08147.

[39] Z. Kasner and O. Dusek. Beyond traditional benchmarks: Analyzing behaviors of open LLMs on data-to-text generation. In L.-W. Ku, A. Martins, and V. Srikumar, editors, *Proceedings of the 62nd Annual Meeting of the Association for Computational Linguistics (Volume 1: Long Papers)*, pages 12045–12072, Bangkok, Thailand, Aug. 2024. Association for Computational Linguistics. doi: 10.18653/v1/2024.acl-long.651. URL https://aclanthology.org/2024.acl-long.651/.

[40] R. Koshkin, K. Sudoh, and S. Nakamura. Transllama: Llm-based simultaneous translation system, 2024. URL https://arxiv.org/abs/2402.04636.

[41] H. Koziolek, S. Grüner, R. Hark, V. Ashiwal, S. Linsbauer, and N. Eskandani. Llm-based and retrieval-augmented control code generation. In *Proceedings of the 1st International Workshop on Large Language Models for Code*, LLM4Code '24, page 22–29, New York, NY, USA, 2024. Association for Computing Machinery. ISBN 9798400705793. doi: 10.1145/3643795.3648384. URL https://doi.org/10.1145/3643795.3648384.

[42] W. Kwon, Z. Li, S. Zhuang, Y. Sheng, L. Zheng, C. H. Yu, J. Gonzalez, H. Zhang, and I. Stoica. Efficient memory management for large language model serving with pagedattention. In *Proceedings of the 29th Symposium on Operating Systems Principles*, SOSP '23, page 611–626, New York, NY, USA, 2023. Association for Computing Machinery. ISBN 9798400702297. doi: 10.1145/3600006.3613165. URL https://doi.org/10.1145/3600006.3613165.

[43] W. Lee, J. Lee, J. Seo, and J. Sim. InfiniGen: Efficient generative inference of large language models with dynamic KV cache management. In *18th USENIX Symposium on Operating Systems Design and Implementation (OSDI 24)*, pages 155–172, Santa Clara, CA, July 2024. USENIX Association. ISBN 978-1-939133-40-3. URL https://www.usenix.org/conference/osdi24/presentation/lee.

[44] C. Li, Y. Sun, L. Jin, L. Xu, Z. Cao, P. Fan, D. Kaeli, S. Ma, Y. Guo, and J. Yang. Priority-based pcie scheduling for multi-tenant multi-gpu systems. *IEEE Computer Architecture Letters*, 18(2):157–160, 2019. doi: 10.1109/LCA.2019.2955119.

[45] K. Li and Y. Zhang. Planning first, question second: An LLM-guided method for controllable question generation. In L.-W. Ku, A. Martins, and V. Srikumar, editors, *Findings of the Association for Computational Linguistics: ACL 2024*, pages 4715–4729, Bangkok, Thailand, Aug. 2024. Association for Computational Linguistics. doi: 10.18653/v1/2024.findings-acl.280. URL https://aclanthology.org/2024.findings-acl.280/.

[46] Z. Li, L. Zheng, Y. Zhong, V. Liu, Y. Sheng, X. Jin, Y. Huang, Z. Chen, H. Zhang, J. E. Gonzalez, and I. Stoica. AlpaServe: Statistical multiplexing with model parallelism for deep learning serving. In *17th USENIX Symposium on Operating Systems Design and Implementation (OSDI 23)*, pages 663–679, Boston, MA, July 2023. USENIX Association. ISBN 978-1-939133-34-2. URL https://www.usenix.org/conference/osdi23/presentation/li-zhouhan.

[47] B. Lin, C. Zhang, T. Peng, H. Zhao, W. Xiao, M. Sun, A. Liu, Z. Zhang, L. Li, X. Qiu, S. Li, Z. Ji, T. Xie, Y. Li, and W. Lin. Infinite-llm: Efficient llm service for long context with distattention and distributed kvcache, 2024. URL https://arxiv.org/abs/2401.02669.

[48] S.-C. Liu, S. Wang, W. Lin, C.-W. Hsiung, Y.-C. Hsieh, Y.-P. Cheng, S.-H. Luo, T. Chang, and J. Zhang. Jarvix: A llm no code platform for tabular data analysis and optimization, 2023. URL https://arxiv.org/abs/2312.02213.

[49] Y. Lu, W. Zhu, L. Li, Y. Qiao, and F. Yuan. Llamax: Scaling linguistic horizons of llm by enhancing translation capabilities beyond 100 languages, 2024. URL https://arxiv.org/abs/2407.05975.

[50] P. Ma, R. Ding, S. Wang, S. Han, and D. Zhang. InsightPilot: An LLM-empowered automated data exploration system. In Y. Feng and E. Lefever, editors, *Proceedings of the 2023 Conference on Empirical Methods in Natural Language Processing: System Demonstrations*, pages 346–352, Singapore, Dec. 2023. Association for Computational Linguistics. doi: 10.18653/v1/2023.emnlp-demo.31. URL https://aclanthology.org/2023.emnlp-demo.31/.

[51] P. Patel, E. Choukse, C. Zhang, A. Shah, Í. Goiri, S. Maleki, and R. Bianchini. Splitwise: Efficient generative llm inference using phase splitting. In *2024 ACM/IEEE 51st Annual International Symposium on Computer Architecture (ISCA)*, pages 118–132, 2024. doi: 10.1109/ISCA59077.2024.00019.

[52] A. Patke, D. Reddy, S. Jha, H. Qiu, C. Pinto, C. Narayanaswami, Z. Kalbarczyk, and R. Iyer. Queue management for slo-oriented large language model serving. In *Proceedings of the 2024 ACM Symposium on Cloud Computing*, SoCC '24, page 18–35, New York, NY, USA, 2024. Association for Computing Machinery. ISBN 9798400712869. doi: 10.1145/3698038.3698523. URL https://doi.org/10.1145/3698038.3698523.

[53] R. Peng, K. Liu, P. Yang, Z. Yuan, and S. Li. Embedding-based retrieval with llm for effective agriculture information extracting from unstructured data, 2023. URL https://arxiv.org/abs/2308.03107.

[54] A. Qiao, S. K. Choe, S. J. Subramanya, W. Neiswanger, Q. Ho, H. Zhang, G. R. Ganger, and E. P. Xing. Pollux: Co-adaptive cluster scheduling for goodput-optimized deep learning. In *15th USENIX Symposium on*